\documentclass[10pt,twoside]{article}
\usepackage{pkas2}
\usepackage{graphicx}
\makehead{M.Karouzos}{LEGACY OF AKARI}

\def\lesssim{\mathrel{\hbox{\rlap{\hbox{\lower4pt\hbox{$\sim$}}}\hbox{$<$}}}}
\def\gtrsim{\mathrel{\hbox{\rlap{\hbox{\lower4pt\hbox{$\sim$}}}\hbox{$>$}}}}

\begin{document}
\sloppy
\pagenumbering{arabic}
\twocolumn[
\pkastitle{27}{1}{2}{2012}
\begin{center}
{\large \bf {\sf
Radio-AGN in the AKARI-NEP field and their role in the evolution of galaxies}}\vskip 0.25cm
{\sc M. Karouzos, M. Im}\\
Department of Physics and Astrnomy, Seoul National University, Seoul 151-747, Korea \\
Center for the Exploration of the Origin of the Universe\\
{\it {E-mail: mkarouzos@astro.snu.ac.kr} }\\
and {\sc the akari-nep team}\\
\normalsize{\it (Received July 1, 2012; Accepted ????)}
\end{center}
\newabstract{Radio-loud active galaxies have been found to exhibit a close connection to galactic mergers and host galaxy star-formation quenching. We present preliminary results of an optical spectroscopic investigation of the AKARI NEP field. We focus on the population of radio-loud AGN and use photometric and spectroscopic information to study both their star-formation and nuclear activity components. Preliminary results show that radio-AGN are associated with early type, massive galaxies with relatively old stellar populations.
\vskip 0.25cm
{\em key words:} infrared - telescope: conferences - proceedings}
\vskip 0.05cm  \flushbottom
]

\newsection{Introduction}

In the context of galaxy evolution in the Universe, the role of nuclear activity, in particular radio-loud active galactic nuclei (AGN), is still under debate. Are radio-AGN a phase of a galaxy's evolution? How are they triggered and what is their effect on their host galaxy? We identify radio-AGN within the AKARI-NEP field and study their host galaxy properties in terms of an hierarchical evolutionary scheme.


\newsection{Cross-identification}
We cross-identify all AKARI-NEP (wide and deep) sources detected in the N2 band of AKARI with the sources from the WSRT catalog at 1.5GHz ([WH10]), following [DO86] (also see [WH12]). In total 401 and 168 radio sources are matched for NEP-wide and -deep, respectively. Photo-z for NEP-deep cross matched sources range between 0.37and 2.2, with most sources having z between 0.37 and 1.\\
We also cross-identify 1.5GHz WSRT sources with the optical spectroscopy catalogs available (Shim et al., Takagi et al., private communication). For a matching radius of 3 arcsec, 48 radio sources are matched (spec-z between 0.03 and 4, with a few above 1). Radio-samples are defined in Table 1.


\begin{table}[!t]
\begin{center}
\scriptsize
\bf{\sc  Table 1.}\\
\label{tab:samples}
\sc{Radio samples selection criteria} \\
\begin{tabular}{ccc}
\\ \hline \hline
Sample & Description & Selection \\
\hline
(1) & All & - \\
(2) & Luminosity & $L_{1.5GHz}>10^{23}$W/Hz* \\
(3) & Flat-spectrum & $\alpha_{radio}<0.5$** \\
\hline
\multicolumn{3}{c}{*limit definition following [CO92][MA07]} \\
\multicolumn{3}{c}{**$\alpha_{radio}$ calculated using 2 or 3 bands} \\
\hline
\end{tabular}
\end{center}
\end{table}

\begin{figure}[!ht]
\resizebox{\hsize}{!}{\includegraphics{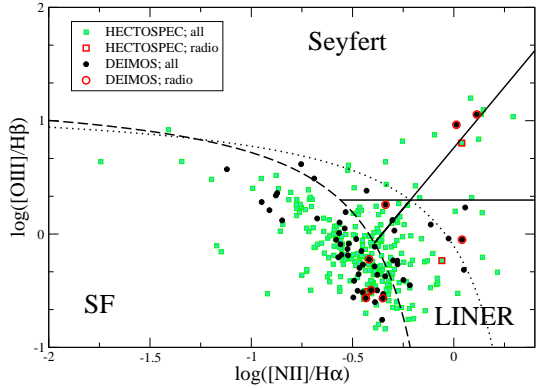}}
\caption{Baldwin-Phillips-Terlevich (BPT) emission line ratios diagnostic diagram to separate star-forming galaxies from active AGN and LINERs. The dashed line is from [n], the dotted line from [l], and the continuous lines are from [m] (horizontal) and [n] (diagonal). They separate star-forming galaxies, LINERs, and Seyferts, respectively. }
\label{fig:bpt}
\end{figure}

\newsection{Optical spectroscopy}
We are in the process of analyzing all the available optical spectra using IDL routines (emission line fluxes, equivalent widths, 4000$\AA$ break, etc.). Using the BPT  emission line classification diagram (Fig. \ref{fig:bpt}) we find 10 AGN/LINERs and 21 transitional objects in a total of 84 sources. 3 radio-sources are classified as AGN/LINERs and 3 as transitional.

\begin{figure}[!ht]
\resizebox{\hsize}{!}{\includegraphics{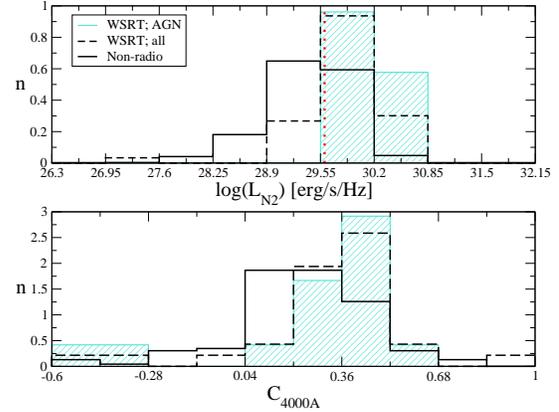}}
\caption{Normalized distributions of the 4000A break index C4000A and rest-frame N2 luminosity for all non-radio sources (black), all radio sources (dashed black), and luminosity-selected radio-AGN (shaded turquoise). Only sources with spectroscopic redshifts are included here. The dotted line denotes the L$_{*}$.}
\label{fig:color}
\end{figure}

\newsection{Optical colors and stellar ages}
We find that luminosity-selected AGN are predominantly associated with elliptical galaxies (u-r$>$2.22;[ST01]), with a similar trend for our flat-spectrum sample. A two-sample KS test gives a 99.8\% probability that sample (2) is drawn from a different population than its parent sample. For the comparison with sample (3) the KS test does not provide a significant result. Both samples (1) and (2) show $C_{4000\AA}$ characteristic of old stellar populations and early-type galaxies (e.g.,[GA05]). A small fraction of sample (2) shows low values of C4000 indicative of a strong power-law non-thermal continuum (Fig. \ref{fig:color}). Assuming that rest-frame N2 luminosity is a good proxy for the stellar mass of a galaxy, both samples (1) and (2) inhabit more massive galaxies compared to the non-radio sample.

\acknowledgments{MK acknowledges the support from the Creative Research Initiative program, No. 2010-0000712, of the National Research Foundation of Korea (NRFK) funded by the Korea government(MEST).}

\end{description}

\end{document}